\documentclass[reprint,aps,longbibliography,amsmath,amssymb,nofootinbib,showpacs,twocolumn]{revtex4-1}
\usepackage{textcomp,overpic}
\usepackage{color}

\usepackage{graphicx}
\usepackage{amsmath,amssymb,epsfig}

\newcommand{\rd}{\mathrm{d}}
\newcommand{\pd}[2]{\frac{\partial #1}{\partial #2}}

\newcommand{\EQE}{{\rm EQE}}

\newcommand{\beqa}{\begin{eqnarray}}
\newcommand{\eeqa}{\end{eqnarray}}
\newcommand{\beqas}{\begin{eqnarray*}}
\newcommand{\eeqas}{\end{eqnarray*}}
\newcommand{\beq}{\begin{equation}}
\newcommand{\eeq}{\end{equation}}

\newcommand{\bal}{\begin{align}}
\newcommand{\ealign}{\end{align}}

\begin{document}

\title{Organic solar cell design as a function of radiative quantum efficiency}
\author{Blaise Godefroid and Gregory Kozyreff}
\email{blaise.godefroid@ulb.ac.be, gkozyref@ulb.ac.be}
\affiliation{Optique Nonlin\'eaire Th\'eorique, Universit\'e libre de Bruxelles (U.L.B.), CP 231, Belgium}
\begin{abstract}
We study the radiative decay, or fluorescence, of excitons in organic solar cells as a function of its geometrical parameters. Contrary to their non-radiative counterpart, fluorescence losses strongly depend on the environment. By properly tuning the thicknesses of the buffer layers between the active regions of the cell and the electrodes, the exciton lifetime and, hence, the exciton diffusion length can be increased. The importance of this phenomenon depends on the radiative quantum efficiency, which is the fraction of the exciton decay that is intrinsically due to fluorescence. Besides this effect, interferences within the cell control the efficiency of sunlight injection into the active layers. An optimal cell design must rely on the consideration of these two aspects. By properly managing fluorescence losses, one can significantly improve the cell performance. To demonstrate this fact, we use realistic material parameters inspired from literature data and obtain an increase of power conversion efficiency from 11.3\% to 12.7\%.  Conversely, not to take into account the strong dependence of fluorescence on the environment may lead to a sub-optimal cell design and a degradation of cell performance. The presence of radiative losses, however small, significantly changes the optimal thicknesses. We illustrate this latter situation with experimental material data.
\end{abstract}
\date{\today}
\maketitle

\section{Introduction} 

Organic solar cells attract considerable research interest as a low-cost, mechanically flexible, and low-temperature manufactured alternative to inorganic solar cells. Various architectures exist, including donor/acceptor bilayer hetero-junction \cite{Tang-1986}, bulk hetero-junction \cite{park2009}, tandem structure \cite{ameri2009}, and cascaded exciton-dissociating hetero-junctions \cite{Cnops-2014}. One of the main factors that limits the efficiency of organic solar cells is the short diffusion length of excitons compared to the absorption length~\cite{Forrest-2005,Kippelen-2009,Mikhnenko-2015}. Indeed, because of the poor transport of photo-generated electrical excitations, the active layers are restricted to thicknesses that are too thin to fully absorb the incident sunlight.  In order to overcome this difficulty, several photonic strategies have been devised to efficiently trap light: randomly structured interface~\cite{Yablonovitch-1982}, hexagonal arrays of nanocolums~\cite{Ko-2009,Liu-2013b}, nanoholes~\cite{Hsiao-2011} or nanospheres~\cite{Grandidier-2011,Yao-2012}. More recently, a photonic fiber plate was shown to significantly improve light trapping through intermittent ray chaos~\cite{Mariano-2014,Mariano-2016}.
 While bulk heterojunctions are an answer to the exciton transport problem, they come with additional difficulties, such as dead-ends in the path of electrons and holes towards their respective electrodes and chemical stability. Moreover, non-radiative recombinations at interfaces between the donor and acceptor are found to severely limit the efficiency of bulk heterojunctions cells~\cite{Kirchartz-2009,Veldman-2008}. Hence, bulk heterojunctions are not a definitive solution to exciton transport and to increase the diffusion length of excitons in planar solar cells remains a critical objective.

Aside from material engineering, it has been pointed out that radiative losses (\textit{i.e.} the diffusion length) can be optically engineered through the geometrical arrangement of the cell when this one is thin and hence present microcavity effects~\cite{Kozyreff-2013}. Indeed, a radiating exciton is electromagnetically equivalent to an oscillating dipole. The optical power emitted by such a dipole can be increased or decreased in the proximity of boundaries, as was experimentally demonstrated by Drexhage~\cite{Drexhage-1974}. A theoretical treatment of this radiation problem was worked out as early as 1909 by Sommerfeld in his study of antennas~\cite{Sommerfeld-1909,Sommerfeld-1949}. More complete accounts followed Drexhage pioneering experiments, both theoretically~\cite{Chance-1978,Lukosz-1980,Neyts-1998,Wasey-2000b} and experimentally~\cite{Tsutsui-1991,Amos-1997}. In high-Q cavities, it is well-known that spontaneous emission of radiation can be strongly suppressed~\cite{Kleppner-1981}. As for solar cells, their are by construction poor cavities, in order to to let as much light in as possible. Nevertheless, it was found that spontaneous emission, \textit{i.e.} radiative losses, can still be significantly reduced. A general rule to promote this effect is to sandwich the photo-active layer by low-index regions. In particular, with $n_1$ and $n_2$, the refractive indexes of the active layer and its surrounding, respectively, the rate of spontaneous emission can be reduced up to a factor $\left(n_2/n_1\right)^5$ for excitons with perpendicular orientation~\cite{Kozyreff-2013}.
In addition to the above aspect, it has been emphasised that a proper choice of layer thicknesses inside the solar cell can significantly influence the distribution of sunlight intensity within the cell and, hence, the absorption of solar photon by the photo-active material~\cite{Peumans-2003,Yoo-2007,Lee-2010,long2011improving,long2011red,Betancur-2012,Salinas-2012,Chen-2014,Chueh-2015}.

The influence of exciton radiative losses on the device performance depends on the radiative quantum efficiency $q$, defined as
\begin{align}
q=\frac{\Gamma_\text{r,bulk}}{\Gamma_\text{bulk}}
&,
&\Gamma_\text{bulk}=\Gamma_\text{r,bulk}+\Gamma_\text{nr},
\end{align}
where $\Gamma_\text{r,bulk}$ and $\Gamma_\text{nr}$ are the rate of radiative and non-radiative decay, respectively, and the `bulk' subscript indicate that it is the bulk value. The factor $q$ is also called `fluorescence quantum efficiency' or `photoluminescence quantum efficiency'. The bulk value of the radiative losses is an intrinsic quantity, which is obtained in the absence of boundary effects. In a confined environment such as in an organic cell, the actual radiative losses $\Gamma_\text{r}(z)$ generally differ from $\Gamma_\text{r,bulk}$ and depend on the position, $z$, of the exciton. The total losses can be written as
\beq
\Gamma(z,q)=\Gamma_\text{bulk}\left(1-q+q\frac{\Gamma_\text{r}(z)}{\Gamma_\text{r,bulk}}\right).
\eeq
In this paper, we seek to optimise the solar cell geometry taking both issues into account: efficient interference for sunlight absorption together with low exciton radiative losses. The fact that it is possible to reduce exciton emission while still efficiently admit light into the solar cell relies on two observations: The exciton emits light (i) at a well-defined frequency, and (ii)  in all directions. Conversely, sunlight comes in a broad spectrum and enters the cell through a narrow cone around the normal direction. In Ref.~\cite{Kozyreff-2013}, only the short-circuit current was evaluated, and a simplified incident spectrum was assumed, which did not allow one to evaluate the cell power conversion efficiency under a realistic illumination.

In order to assess device performance, we will first establish a generalisation of the Shockley-Queisser (S-Q) theory~\cite{Shockley-1961} for organic solar cells. Indeed, in this classic theory, the transport of electrical excitations is not an issue, since only inorganic semi-conductor materials with high charge mobilities are considered. Besides, microcavity effects are absent from  S-Q theory. We will therefore generalise S-Q theory to take into account the diffusive exciton transport and microcavity effects both for sunlight injection and exciton radiation. In this more general theory, the External Quantum Efficiency (EQE), defined as the number of electrons generated per incoming photons, essentially replaces the material absorptivity.

The rest of the paper is  organised as follows. In Sec.~\ref{sec:SQ}, we present the generalised S-Q theory, taking into account exciton transport with space-dependent  radiative losses. In  Sec.~\ref{sec:num}, we apply our theory to several cell geometries and discuss the impact of $q$ on the optimal cell design. Finally, we conclude.

\section{Generalized Shockley-Queisser theory}\label{sec:SQ}

\subsection{Shockly-Queisser detailed balance theory for conventional solar cells}
In its simplest form, S-Q theory of solar cells derives from the following statement of detailed balance:
\begin{equation}
I_s=I_R(V)+I(V).\label{db1}
\end{equation} 
Above, $I_s$ is the number of electron-hole (e-h) pairs generated per unit of time and area by photon absorption. From this current, a portion $I_R(V)$ will recombine to produce a thermal electromagnetic radiation, leaving a particle current $I(V)$ (hence an electrical current $-eI(V)$) of electrons to flow in the external circuit under an electrical potential $V$. For the sake of the present discussion, non-radiative recombination processes are omitted from the right hand side of Eq.~(\ref{db1}).

If each absorbed photon leads to the generation of one and only one electron-hole pair then, 
\beq
I_{s}=\Omega_s\int_{0}^\infty a(\lambda)\phi_{AM1.5}(\lambda) \rd\lambda
\eeq
where $\Omega_s=6.85 \times 10^{-5}$ is the solid angle under which illumination is received from the sun, $a(\lambda)$ is the absorptivity, $\phi_{AM1.5}$ is the AM1.5 solar spectrum (in photons s$^{-1}$m$^{-2}$m$^{-1}$sr$^{-1}$), and $\lambda$ is the optical wavelength. We restrict our attention to illumination at normal incidence. Similarly, the recombination term is given by
\begin{align}
I_{R }(V)&=2\pi\int_0^{\frac\pi2}\sin\theta\cos\theta  \int_{0}^\infty a(\lambda)\phi(\lambda,T,V) \rd\lambda\rd\theta, \label{IR}\\
\phi(\lambda,T,V) &= \left(2 c/\lambda^4\right)\left[\exp \left(\frac{hc/\lambda-eV}{k T} \right)-1\right]^{-1}. \label{eq:Planck} 
\end{align}
Above, $\phi(\lambda,T,V)$ is the Planck's distribution in which the chemical potential of radiation is given by $eV$ \cite{Wurfel-1982,Nelson, DeVos-2008},  $c$ is the speed of light, $h$ is Planck's constant, $k$ is Boltzmann's constant, and $T$ is the cell temperature. In Eq.~(\ref{IR}), we have used Kirchhoff's law and equated the emissivity of the cell with its absorptivity. Thus, Eq.~(\ref{db1}) becomes
\begin{equation}
I(V)= \int_{0}^\infty a(\lambda)\left[\Omega_s\phi_{AM1.5}(\lambda)  - \pi\phi(\lambda,T,qV)\right]\rd\lambda. \label{SQ2}
\end{equation} 
This last equation eventually leads to the Shockley-Queisser's limit  if we assume that $a(\lambda)$ is a step function. Stated in the above way, S-Q theory can simply be generalised to more general cell, such as those in which the internal electrical transport is controlled by excitons.

\subsection{Exciton-regulated solar cell}
In organic solar cells the absorption of a photon gives rise to an exciton, with a binding energy that is large compared to $kT$. The exciton are therefore long-lived; moreover, being electrically neutral, their motion is governed by diffusion. It is only at the interface between the donor and acceptor materials that the local electric gradient can break the exciton into a free hole and a free electron. Neglecting other processes, elecron-hole pairs are only generated by exciton dissociation at the donor/acceptor interface and they subsequently disappear only by radiative recombination or by being collected in the external circuit. Thus, 
the detailed balance equation~(\ref{db1}) becomes
\begin{equation}
I_\chi(\phi_{AM1.5})=I(V)+I_R(V)\label{db2}
\end{equation}
where $I_\chi(\phi_{AM1.5})$ is the exciton dissociation current under the AM1.5 illumination equal to the number of e-h pair generated per unit of time and area. This term is independent of voltage since excitons are neutral particles. It can be expressed as
\begin{equation}
I_\chi(\phi_{AM1.5})=\Omega_s\int_{0}^\infty \EQE(\lambda,0)\phi_{AM1.5}(\lambda)\rd\lambda
\end{equation}
where we define $\EQE(\lambda,\theta)$ as the number of e-h pair generated per incident photon with wavelength $\lambda$ and incidence angle $\theta$ with respect to the normal direction. This coincides with the usual definition if the charge collection efficiency is unity at short-circuit, which is a common assumption~\cite{Peumans-2003,Yoo-2004}.

Next, we must establish the radiative current $I_R(V)$. We first note that, in the dark, under an ambient isotropic illumination, $\phi(\lambda,T,0)$, it must be equal to the e-h generation current, $I_\chi$, as equilibrium conditions request. 
\beq
I_R(0)=2\pi\int_0^{\frac\pi 2}\sin\theta\cos\theta \int_{0}^\infty \EQE(\lambda,\theta)\phi(\lambda,T_c,0) \rd\lambda\rd\theta
\label{eq:V=0}
\eeq
Outside equilibrium, the cell emits light with a potential of radiation as in Eq.(\ref{eq:Planck}). In that case, one postulates that Eq.~(\ref{eq:V=0}) can be extrapolated to non-zero values of $V$ as
\begin{equation}
I_R(V)=2\pi\int_0^{\frac\pi 2}\sin\theta\cos\theta \int_{0}^\infty \EQE(\lambda,\theta)\phi(\lambda,T_c,V) \rd\lambda\rd\theta\label{IR2}
\end{equation}
The validity of this modelling step is discussed and confirmed in~\cite{Kirchartz-2016}. Note that this last expression is similar to Eq.~(\ref{IR}) but with the absorptivity, $a(\lambda)$, replaced by  $\EQE(\lambda,\theta)$~\cite{Vandewal-2012}. Finally, Eq.~(\ref{db2}) yields 
\begin{multline}
I(V)=\Omega_s\int_{0}^\infty \EQE(\lambda,0)\phi_{AM1.5}(\lambda)\rd\lambda\\
- 2\pi\int_0^{\frac\pi 2}\sin\theta\cos\theta \int_{0}^\infty \EQE(\lambda,\theta)\phi(\lambda,T_c,V) \rd\lambda\rd\theta\label{SQ3}
\end{multline}
A similar derivation can be found in~\cite{Rau-2007,Rau-2017}. Other phenomenological factors such as series and parallel resistances and the ideality factor are unnecessary for the present discussion and are therefore omitted from the model. Note that the theory so far is very general and is valid beyond the theory of organic cells.  In order to determine the current-voltage curve (\ref{SQ3}) for a given cell, one has to calculate $\EQE(\lambda,\theta)$. This is what we do in the next section for bilayer heterojunction organic solar cells.

\subsection{Calculation of EQE}
We now wish to compute $\EQE(\lambda,\theta)$, \textit{i.e.} the number of e-h pairs produced at the Donor/Acceptor interface  per incoming photon as a function of the photon wavelength and incidence angle.

Let $N(\lambda,\theta)=\phi_{AM1.5}(\lambda)\cos\theta$ be the number of photons that are incident per unit time and device area at an angle $\theta$. This results in an electromagnetic intensity distribution $g(z,\lambda,\theta)$ inside the device, which can be computed by  the transfer-matrix method~\cite{BornWolf,Pettersson-1999} (unpolarised light assumed.)  With the proper normalisation for $g(z,\lambda,\theta)$ that takes into account the absorption coefficient and the conversion efficiency of absorbed photons into excitons, the rate of production of excitons per unit length is $N(\lambda,\theta)g(z,\lambda,\theta)$. Hence the distribution of excitons $\rho$ produced by the flux $N(\lambda,\theta)$ satisfies
\beq
D \pd{^2\rho}{z^2}-\Gamma(z,q)\rho+N(\lambda,\theta)g(z,\lambda,\theta)=0,
\label{eq:dimensional}
\eeq
with
\begin{subequations}
\begin{align}
\pd{\rho}{z}&=0, &z=z_{-1},\\
\rho&=0, &z=z_{0},\\
\pd{\rho}{z}&=0, &z=z_{1}.
\end{align}\label{eq:bc}
\end{subequations}
\hspace*{-.17cm}Above, $D$ is the exciton diffusion constant, which takes the value $D_A$ or $D_D$ in the acceptor or donor material, respectively. Next,  $\Gamma(z,q)$ is  the exciton decay rate, which is computed as in~\cite{Kozyreff-2013} assuming random exciton orientation. Finally, Eqs.~(\ref{eq:bc}) express a no-flux boundary condition for excitons at the interfaces ($z=z_{\pm1}$) between the active layers and the adjacent blocking layers, while $\rho(z_0)=0$ models complete excitons dissociation into free electrons and holes at the Donor/Acceptor interface~\cite{Peumans-2003,Yoo-2004,Scully-2006}. Exciton dissociation into free electrons and holes is assumed negligible anywhere else \cite{stubinger-2001}.

Note that if we divide Eq.~(\ref{eq:dimensional}) by $\Gamma_\text{bulk}$ and normalise $\rho$ as $\rho=\left(N/\Gamma_\text{bulk}\right)\rho'$, the exciton transport equation becomes
\begin{align}
L^2 \pd{^2\rho'}{z^2}-\frac{\Gamma(z,q)}{\Gamma_\text{bulk}}\rho'+g(z,\lambda,\theta)&=0,
&L^2=\frac{D}{\Gamma_{bulk}}.
\label{eq:diff}
\end{align}
Solving that equation, the diffusive current at the Donor/Acceptor interface yields the EQE  as
\begin{multline}
\text{\EQE}(\lambda,\theta)=\frac{1}{N(\lambda)}\left(\left. D_{ D}\pd{\rho}{z}\right|_{z_0+\epsilon}-\left. D_{A}\pd{\rho}{z}\right|_{z_0-\epsilon}\right)\\
=
 L^2_{ D} \left.\pd{\rho'}{z}\right|_{z_0+\epsilon}- L^2_{A}\left.\pd{\rho'}{z}\right|_{z_0-\epsilon}
\label{EQE1}.
\end{multline}
Once, $\EQE(\lambda,\theta)$ is determined, the $I(V)$ curve and the power conversion efficiency can be computed from Eq.~(\ref{SQ3}).

\section{Numerical results}\label{sec:num}
\begin{figure*}
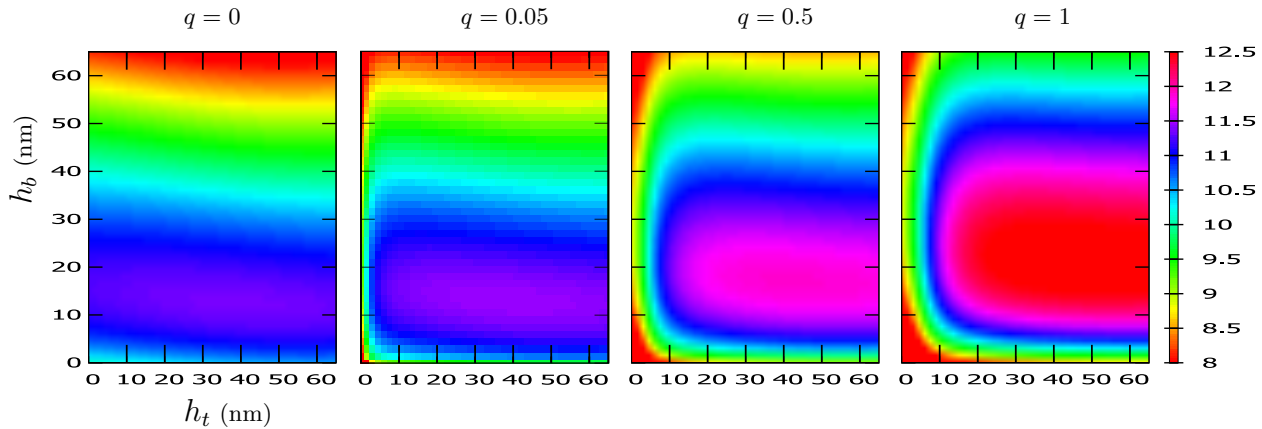

\begin{overpic}[ width=16.cm,height=4.5cm]{greg} 
\put(11,30){$q=0$}
\put(34,30){$q=0.05$}
\put(57,30){$q=0.5$}
\put(80,30){$q=1$}
\put(-3.5,15){\rotatebox{90}{{\large$h_b$} (nm)}}
\put(11,-3){{\large$h_t$} (nm)}
\end{overpic}
\vspace{.3cm}
\caption{(Color online) Power conversion efficiency $\eta$ for the configuration presented in table \ref{table1} as a function of blocking layer thicknesses, $h_t$ and $h_b$, for $q=0, 0.05, 0.5, 1$ in both Acceptor and Donor layers.}
\label{figideal}
\end{figure*}

In this section, we  illustrate our theory with two  bi-layer cells, each one of the form Ag/HBL($h_b$)/Acceptor/Donor/EBL($h_t$)/ITO/glass. Here, Ag designates the back silver electrode, HBL is a hole-blocking and electron-transporting layer, with thickness $h_b$,  and EBL is an electron-blocking and hole-transporting layer, with thickness $h_t$.  Finally, ITO designates a transparent indium tin oxide electrode (ultra-thin metal film could be considered instead \cite{Chueh-2015,Huang-2015} but we omit this possibility here, as this does not change the general conclusions.)

Given the large number of geometrical parameters available, we choose to maximise the cell power conversion efficiency 
by varying only the thickness $h_b$ and $h_t$ of the HBL and EBL, respectively, for given $q$.   
These two thicknesses are easily tunable fabrication parameters and it has been pointed out that they can significantly affect the efficiency of the cell \cite{Peumans-2003,Yoo-2007,Lee-2010,long2011improving,long2011red,Betancur-2012,Salinas-2012,Chen-2014,Chueh-2015}.

As we vary $q$, we keep the bulk diffusion length unchanged, so as to compare active regions that would otherwise be equivalent. The power conversion efficiency is given by
\beq
\eta= \max_V\left[-eI(V)V\right]\bigg/\Omega_s\int_0^{\infty}\left(hc/\lambda\right)\phi_{AM1.5}(\lambda)\rd\lambda.
\eeq
We consider two distinct situations. First, we demonstrate the potential benefit from using large-$q$ photoactive organic molecules, if the cell is properly designed. To this end, we assume a favourable set of refractive indexes and thicknesses within ranges that are dictated by the literature. We show that a substantial gain in cell efficiency can be obtained. 

Secondly, we simulate the following cell: Al/BCP/C$_{70}$/DBP/MoO$_3$/ITO/glass, for which we use experimentally measured refractive indexes\footnote{Courtesy of Jordi Martorell and Marina Mariano Juste}. Here, BCP stands for Bathocuproine and DBP for Tetraphenyldibenzoperiflanthene. The photovoltaic junction is achieved by C$_{70}$ (Acceptor) and DBP (Donor). BCP and and MoO$_3$ are electron and hole-transporting layer, respectively. In the study of this latter cell, we simulate the possibility that C$_{70}$ and DBP would be replaced by equivalent molecules, C$_{70}^*$ and DBP$^*$, with identical complex refractive indexes and bulk diffusion lengths, but variable radiative quantum efficiency $q$. For this cell, the refractive indexes do not yield an increase of $\eta$ with $q$. However, the optimal BCP and MoO$_3$ thicknesses do vary substantially with $q$. 

In our computation of $\Gamma(z,q)$, we assumed that radiative excitons emit at 900nm. Further the distribution $g(z,\lambda,\theta)$ in Eq.~(\ref{eq:diff}) was computed by assuming that the incoming field was an even combination of TE and TM waves at oblique incidence. Finally,  we assume  the same value of $q$ for the two active materials.

\subsection{First example: high-efficiency cell}\label{num}

\begin{table}[b]
\caption{\label{table1}Optical parameters for an example cell. Imaginary part of refractive index in acceptor and donor layer non-zero only for $\lambda\in$~[300nm,7000nm]} 
\begin{ruledtabular}
\begin{tabular}{cccc}
 Layer & Thickness & Refractive index & Diffusion length \\ 
\hline glass & $\infty$ & 1.45 & \\
ITO & 150 nm & 1.76+0.08i & \\
EBL & $h_t$ & 1.7 &\\
Donor & 15 nm & 2.8+0.85i & 10 nm\\
Acceptor &  15 nm &  2.8+0.85i & 10 nm\\
HBL &  $h_b$ &  1.7 &\\
Ag &  $\infty$ &  0.03+ 5.19i&\\
\end{tabular}
\end{ruledtabular}
\end{table}

We first consider a favourable set of refractive indexes for increasing the diffusion length through the management of radiative losses. This set, together with layer thicknesses, is given in Table~\ref{table1}. All refractive indexes are assumed independent of wavelength, except the imaginary part in the active layers, which is non-zero only between 300 and 700 nm. The assumed values of the refractive indexes for the blocking layers are consistent with the literature~\cite{Peumans-2003,Rand-2007,Chen-2014} and with the index of BCP. The large index in the active region is chosen to maximise the contrast between the blocking and the photo-active layers, in accordance with the conclusion in Ref.~\cite{Kozyreff-2013}. A larger ratio of the real part of the refractive index in the photo-active layer and the blocking layers is liable to improve the cell efficiency. Furthermore, we assume an exciton  diffusion length of 10~nm in both donor and acceptor layer, while the thickness of these layers is chosen to be  $1.5$ times the diffusion length. The ITO thickness of 150~nm is chosen so as to promote good light injection in the active layers thanks to constructive interferences in the spectral range of interest. To avoid spurious resonances in the glass capping, we assume this layer to have  infinite thickness and correct the incoming intensity accordingly.

\begin{figure}[h]
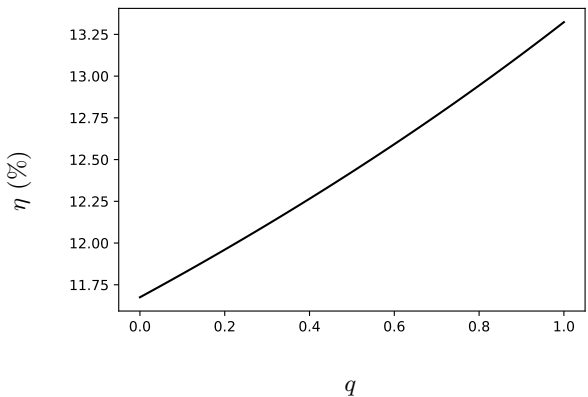

\begin{overpic}[width=8.0cm]{nq} 
\put(-6,25){\rotatebox{90}{$\eta \,\, (\%)$}}
\put(50,-5){$q$}
\end{overpic}
\vspace{.3cm}
\caption{Power conversion efficiency $\eta$ as a function of radiative quantum efficiency $q$, for the parameters given in table \ref{table1} and $(h_t,h_b)=(45\rm{nm},21\rm{nm})$.}\label{fign}
\end{figure}

\begin{figure}[h]
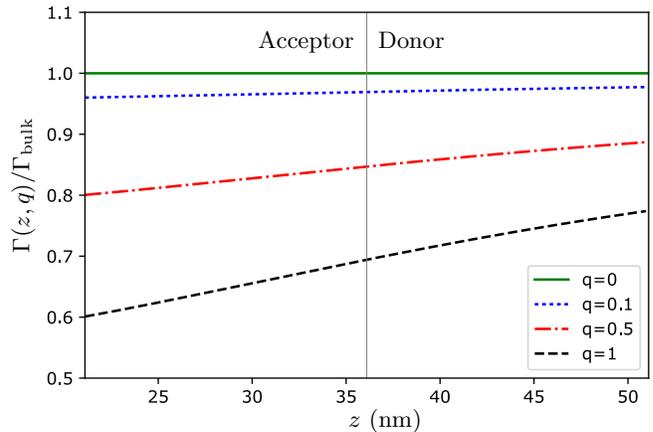

\begin{center}
\begin{overpic}[ width=8.cm]{bgreg} 
\put(-6,25){\rotatebox{90}{$\Gamma(z,q)/\Gamma_\text{bulk}$}}
\put(50,-4){$z$ (nm)}
\put(35,59){Acceptor}
\put(55,59){Donor}
\end{overpic}
\caption{(Color online) Normalised exciton decay rate as a function of the distance from the back electrode for the configuration presented in table \ref{table1} with $(h_t,h_b)=(45\rm{nm},21\rm{nm})$. The vertical line separates the Donor and Acceptor layers.}
\label{figb2}
\end{center}
\end{figure}

Figure~\ref{figideal} shows the cell efficiency as a function of the top and bottom blocking layer thicknesses, $h_t$ and $h_b$ for three values of $q$. One notices that there is a qualitative change in the graph as soon as $q>0$, with low efficiency for very small values of either $h_t$ or $h_b$. This is due to the quenching effect experienced by radiative excitons in the vicinity of a dissipative medium. As $q$ progresses from $q=0$ to $q=1$, the optimal geometry vary from  $(h_t,h_b)$=(40\,nm, 12\,nm) to $(h_t,h_b)$=(45\,nm, 21\,nm). Meanwhile the cell efficiency steadily progresses from $11.3\%$ to $12.7\%$, see Fig.~\ref{fign}. For the most efficient configuration, we plot in Fig.~\ref{figb2}, the space-dependent decay rate $\Gamma(z,q)$ for different values of $q$. In that figure, the advantage of managing the exciton radiative decay appears clearly.

\subsection{Second example: low efficiency cell }\label{num}
\begin{figure*}
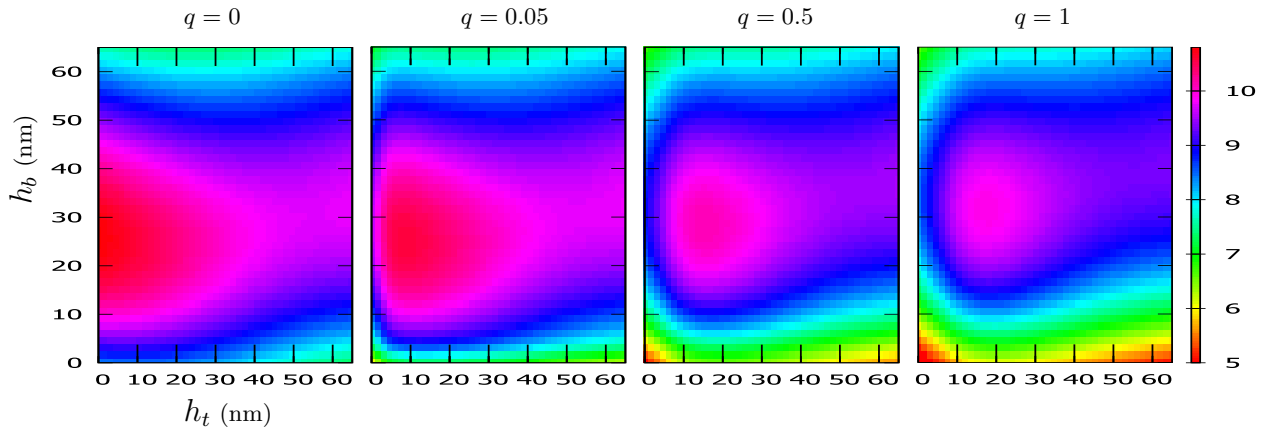

\begin{center}
\begin{overpic}[width=16cm,height=4.5cm]{marina} 
\put(11,30){$q=0$}
\put(34,30){$q=0.05$}
\put(57,30){$q=0.5$}
\put(80,30){$q=1$}
\put(-3.5,15){\rotatebox{90}{{\large$h_b$} (nm)}}
\put(11,-3){{\large$h_t$} (nm)}
\end{overpic}
\vspace{.3cm}
\caption{(Color online) Power conversion efficiency $\eta$ for Al/ BCP($h_b$)/ C$_{70}^*$(31.5 nm)/ DBP$^*$(10.5 nm)/ MoO$_3$($h_t$)/ ITO(150 nm)/ glass as a function of $h_t$ and $h_b$ for $q=0, 0.05, 0.5, 1$ in both Acceptor and Donor layers.  }\label{fig:hot}
\end{center}
\end{figure*}

\begin{figure}[h]
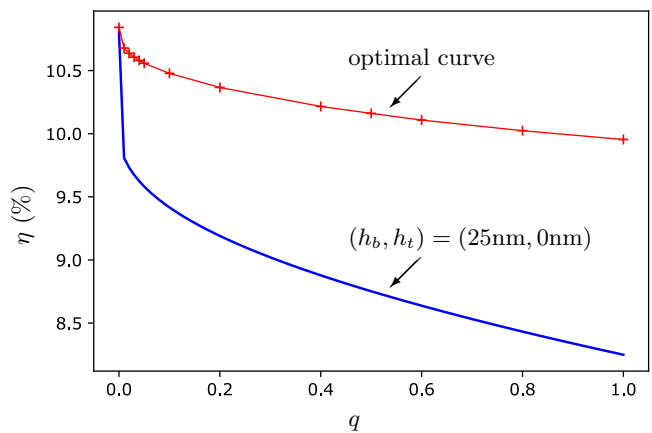

\begin{overpic}[width=8.0cm]{nqmarina} 
\put(-6,25){\rotatebox{90}{$\eta \,\, (\%)$}}
\put(50,-5){$q$}
\put(50,55){optimal curve}
\put(62,53){\vector(-1,-1){5}}
\put(50,25){$(h_b,h_t)=(25\rm{nm},0\rm{nm})$}
\put(62,23){\vector(-1,-1){5}}
\end{overpic}
\vspace{.3cm}
\caption{(Color online) Power conversion efficiency $\eta$ as a function of radiative quantum efficiency $q$, for Al/ BCP($h_b$)/ C$_{70}^*$(31.5\,nm)/ DBP$^*$(10.5\,nm)/ MoO$_3$($h_t$)/ ITO(150\,nm)/ glass cell. Blue curve:  $(h_b,h_t)=(25\rm{nm},0\rm{nm})$ for all $q$. Red curve: $(h_b,h_t)$ set to optimal value for each $q$.}\label{fig:n(q)marina}
\end{figure}

\begin{figure}[h]
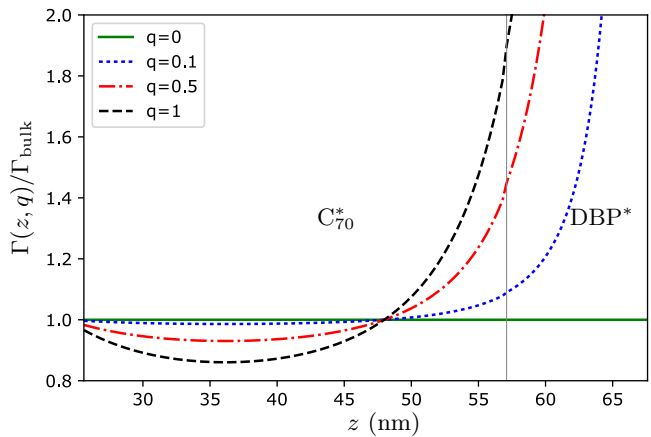

\begin{center}
\begin{overpic}[ width=8.cm]{bmarina} 
\put(-6,25){\rotatebox{90}{$\Gamma(z,q) / \Gamma_\text{bulk}$}}
\put(50,-4){$z$ (nm)}
\put(45,30){C$_{70}^*$}
\put(87,30){DBP$^*$}
\end{overpic}
\vspace{.2cm}
\caption{ (Color online) Normalised exciton decay rate in the active layers for different value of $q$ for the following configuration: Al/ BCP(25\,nm)/ C$_{70}^*$(31.5\,nm)/ DBP$^*$(10.5\,nm)/ MoO$_3$(0\,nm)/ ITO(150\,nm)/ glass. The vertical line separates the two active layers. }
\label{figb1}
\end{center}
\end{figure}

In this second example, we study the following cell, 
Al/ BCP($h_b$)/ C$_{70}^*$(31.5nm)/ DBP$^*$(10.5nm)/ MoO$_3$($h_t$)/ ITO(150nm)/ glass,
where  C$_{70}^*$ and  DBP$^*$ designate materials with identical complex refractive indexes to those of C$_{70}$ and  DBP, but where $q$ is a free parameter. According to Ref.~\cite{Yokoyama-2012}, the diffusion length  is 21nm in C$_{70}$ and 7nm in DBP.  As in the previous example, the C$_{70}^*$ and DBP$^*$ thicknesses are chosen equal to $1.5$ times the diffusion length, consistently with~\cite{Siegmund-2017}.

The graph of $\eta$ as a function of $h_b$ and $h_t$ is given in Fig.~\ref{fig:hot} for four values of $q$. This time the maximum efficiency, 10.8\%, is obtained for $q=0$. The cell performance steadily degrades with increasing $q$,  and do not exceed 10\% for $q=1$. However, it is important to note that the optimal configuration vary significantly between these extreme case: For $q=0$, the best set is near of $(h_b,h_t)$=(25\,nm, 0\,nm); next, for $q=0.05$ the optimal configuration is $(h_b,h_t)$=(25\,nm, 9\,nm);  for $q=0.5$ the optimal configuration is $(h_b,h_t)$=(29\,nm, 15\,nm); finally,  for $q=1$, it becomes $(h_b,h_t)$=(32\,nm, 18\,nm). 

This last observation is a warning sign that the cell architecture should be designed with proper account of $q$. Indeed, with $(h_b,h_t)=(25\,nm,0\,nm)$, which is optimal for $q=0$, $\eta$ drops from 10.8\% to 8.7\% if $q=0.5$ and is only 8.2\% for $q=1$, see Fig~\ref{fig:n(q)marina}. The cause of this rapid decay can be understood from Fig.~\ref{figb1}, which shows $\Gamma(z,q)$ for this configuration and various values of $q$. Here, boundary effects strongly increase the radiative losses in DBP$^*$. While a reduction of radiative losses is achieved in C$_{70}^*$, this is not sufficient to counterbalance the adverse effect in DBP. 

Looking again at Fig.~\ref{fig:n(q)marina}, it is important to notice that there is a sudden drop of $\eta$ as soon as $q$ differs from zero, even if one keeps track of the best possible configuration $(h_b,h_t)$ while varying $q$. This shows that to model the organic cell with $q=0$ is an inaccurate modelling assumption.

\section{Discussion}
In this paper, we have emphasised the importance of properly taking into account the space-dependent rate of radiative decay of exciton, $\Gamma(z,q)$ in organic solar cells. It is well-known that, as soon as the radiative quantum yield $q$ is not zero, the radiative decay rate diverges as an exciton comes into contact with a dissipative surface~\cite{Chance-1978}, leading to exciton quenching. However, it has so far been overlooked that radiative decay can be reduced elsewhere in the cell. Here, we showed that with a proper choice of spacer thicknesses, $h_b$ and $h_t$, this effects leads to a significant increase of the diffusion length and, hence, of the cell efficiency $\eta$. A general rule to exploit this effect is that the (real part of) refractive index contrast between the photoactive layers and the electron- and hole-blocking layers should be large. With a ratio of 2.8/1.7 and random exciton orientation, we have numerically demonstrated an increase of $\eta$ from 11.3\% in the best configuration for $q=0$ to 12.7\% in the best configuration for $q=1$. Note that the gain in efficiency  rapidly increases with the refractive index in the active layers. If we suppose, for instance, that the blocking layers only has a refractive index of 1.45, then the maximum efficiency would increase from 11.3\%  with $q=0$ to 14.3\%  with $q=1$. Conversely, one may seek for active materials with larger refractive indexes. As an example, phtalocyanine (Pc) and its derivatives (subPc, Fluorinated-Pc, Mg-Pc,\ldots), or subnaphtalocyanine (subNc) can display  refractive indexes above 3 and a large radiative quantum yield~\cite{Gommans-2007,Wojdyla-2008,Lunt-2009,Verreet-2009,Verreet-2011,Menke-2013}. Equally, perovkites are found to display large refractive indexes in their absorption range~\cite{Lin-2015} together with high photoluminescence efficiency~\cite{Deschler-2014}.  Another way to drastically improve this gain would be to orient the exciton dipolar moment preferably  along the cell axis, as this can dramatically decrease $\Gamma(z,q)$~\cite{Kozyreff-2013}. 

Based on the above calculation, it would be desirable to develop organic solar cells with molecules having a large  $q$. In this regard, one may turn one's attention to organic light emitting devices (OLED) molecules. With OLEDs, substantial development have already been made to taylor the dipolar emission of excitons~\cite{Flammich-2009,Flammich-2010,Penninck-2012}. Moreover, $q$ of  nearly unity and good exciton orientation has been demonstrated~\cite{Kim-2013}. However, contrary to maximising exciton emission, as in OLEDs, here we want to suppress it. The maximisation of $\eta$ is found to require large spacers, both in order to maximise light injection and to suppress exciton decay. In practice, one is limited by the finite conductivity of these spacers. Nevertheless, large spacer values have been used in OLED~\cite{Penninck-2012} and also considered before in organic solar cells~\cite{Lee-2010,Lassiter-2011,Betancur-2012,Chen-2014,Sahdan-2016}.

It has been claimed that a good solar cell must also be a good emitter~\cite{Miller-2013}. Our conclusion above that solar cells can be improved based on good exciton radiative properties is consistent with this statement. However, it should be stressed that having a large value of $q$ is not sufficient in itself to improve the cell efficiency. This is what we demonstrated in our second numerical example. For refractive indexes corresponding to  Al/ BCP/ C$_{70}$/ DBP/ MoO$_3$/ ITO/ glass, a large value of $q$ tends to degrade the cell performance, because radiative exciton decay is overall increased in the optimal configuration. Still, taking $q$ into account appears crucial. Indeed, if not properly managed, radiative losses can be worse than anticipated. Thus, even in that case, $q$ is an important parameter to take into account.

\subsubsection*{Acknowledgement}
We thank Jordi Martorell and Marina Mariano Juste (ICFO, The Institute of Photonic Sciences) for helpful discussions and for communicating their data. G.K. is a research associate of the Fonds de la Recherche Scientifique -FNRS (Belgium) \\


\end{document}